\documentclass[sigconf]{acmart}

\usepackage{pifont}
\usepackage{booktabs} 
\usepackage{multirow}
\usepackage{multicol}
\usepackage{graphicx}
\usepackage{threeparttable} 
\usepackage[hyphenbreaks]{breakurl}
\usepackage{microtype}
\usepackage[utf8]{inputenc} 
\usepackage{balance}
\usepackage{siunitx}
\usepackage{tcolorbox}
\usepackage{hyperref}

\usepackage{array}
\newcolumntype{P}[1]{>{\centering\arraybackslash}p{#1}}
\newcolumntype{L}[1]{>{\raggedright\let\newline\\\arraybackslash\hspace{0pt}}m{#1}}
\newcolumntype{C}[1]{>{\centering\let\newline\\\arraybackslash\hspace{0pt}}m{#1}}
\newcolumntype{R}[1]{>{\raggedleft\let\newline\\\arraybackslash\hspace{0pt}}m{#1}}

\usepackage{etoolbox}                
\AtBeginEnvironment{tabular}{\footnotesize}

\AtBeginDocument{%
  \providecommand\BibTeX{{%
    \normalfont B\kern-0.5em{\scshape i\kern-0.25em b}\kern-0.8em\TeX}}}

\newtcolorbox{summarybox}[1][]
{
    fontupper=\sffamily, 
    sharp corners,
    left=1mm,
    right=1mm,
    top=1mm,
    bottom=1mm,
    boxrule=0.2mm,
    colback=gray!10!white,
    #1,
}

\acmConference[ICSE 2024]{46th International Conference on Software Engineering}{April 2024}{Lisbon, Portugal}
\copyrightyear{2024}
\acmYear{2024}
\setcopyright{acmlicensed}\acmConference[ICSE '24]{2024 IEEE/ACM 46th International Conference on Software Engineering}{April 14--20, 2024}{Lisbon, Portugal}
\acmBooktitle{2024 IEEE/ACM 46th International Conference on Software Engineering (ICSE '24), April 14--20, 2024, Lisbon, Portugal}
\acmDOI{10.1145/3597503.3639197}
\acmISBN{979-8-4007-0217-4/24/04}

\begin{document}

\title[How Are Paid and Volunteer Open Source Developers Different?]{How Are Paid and Volunteer Open Source Developers Different? A Study of the Rust Project}

\author{Yuxia Zhang}
\affiliation{
  \institution{Beijing Institute of Technology}
  \institution{School of Computer Science \& Technology}
  \city{Beijing}
  \country{China}
}
\email{yuxiazh@bit.edu.cn}

\author{Mian Qin}
\affiliation{
  \institution{Beijing Institute of Technology}
  \institution{School of Computer Science \& Technology}
  \city{Beijing}
  \country{China}
}
\email{qinmian@bit.edu.cn}

\author{Klaas-Jan Stol}
\affiliation{
  \institution{University College Cork and Lero}
  \institution{School of Computer Science and IT}
  \city{Cork}
  \country{Ireland}
}
\email{k.stol@ucc.ie}

\author{Minghui Zhou}
\affiliation{
    \institution{Peking University}
    \institution{School of Computer Science}
    \city{Beijing}
    \country{China}
}
\email{zhmh@pku.edu.cn}

\author{Hui Liu}\authornote{Corresponding author}
\affiliation{%
  \institution{Beijing Institute of Technology}
  \institution{School of Computer Science \& Technology}
  \city{Beijing}
  \country{China}
}
\email{liuhui08@bit.edu.cn}

\renewcommand{\shortauthors}{Zhang et al.}

\begin{abstract}
It is now commonplace for organizations to pay developers to work on specific open source software (OSS) projects to pursue their business goals. 
Such paid developers work alongside voluntary contributors, but given the different motivations of these two groups of developers, conflict may arise, which may pose a threat to a project's sustainability.
This paper presents an empirical study of paid developers and volunteers in Rust, a popular open source programming language project. Rust is a particularly interesting case given considerable concerns about corporate participation. 
We compare volunteers and paid developers through contribution characteristics and long-term participation, and solicit volunteers' perceptions on paid developers. We find that core paid developers tend to contribute more frequently; commits contributed by one-time paid developers have bigger sizes; peripheral paid developers implement more features; and being paid plays a positive role in becoming a long-term contributor.
We also find that volunteers do have some prejudices against paid developers. 
This study suggests that the dichotomous view of paid vs. volunteer developers is too simplistic and that further subgroups can be identified.  
Companies should become more sensitive to how they engage with OSS communities, in certain ways as suggested by this study. 
\end{abstract}


\maketitle

\section{Introduction}\label{sec: intro}
Open source software (OSS) has become part of society's fundamental infrastructure.
A recent report \cite{ossra2022} estimates that up to 97\% of codebases contain OSS. Motivated by the huge impact of OSS, an increasing number of companies have embraced open source to accelerate new innovations, approaches, and best practices \cite{james16, hauge2010adoption}. 
For example, most work in the Linux kernel today is done by paid developers who are hired by companies, emphasizing the important role of paid developers in OSS \cite{corbet2012linux, linux2022companies}. 

Companies participate in OSS projects by tasking their employees to make specific contributions to an OSS project, or by hiring volunteers from the project to do the same. 
Companies participate in OSS to pursue their business objectives \cite{Zhou2016Inflow, zhang2021companies}, which may affect the direction and scale of how paid developers make contributions to OSS projects. 
(We use the term `paid developer' to refer to those who are paid and tasked by a company to make contributions to a specific OSS.)  
Volunteers contribute to open source projects for other reasons, frequently for intrinsic motivations, such as fun, but also to make a better product (to \textit{``scratch an itch''} \cite{raymond1999cathedral}), rather than pursuing commercial objectives. 

Corporate participation in open source is also a source of some concern for several reasons. 
Contributions from a company may be highly specialized and specific to the company, and be at odds with the OSS project's roadmap; 
integrating such contributions into the main development branch may not be welcome \cite{berdou2006insiders}. Recent work found that dominant corporate participation may be a threat to a project's survival rate \cite{zhang2022corporate}. 
Diverging motivations and behavior that do not comply with a project's norms may lead to conflict within a project. Such conflicts may lead to companies dropping their support of an OSS project, but also of individual developers, whether they are paid or volunteers, leaving a community \cite{resigned2021, schaarschmidt2018company}. 

Corporate involvement has also led to concerns in the Rust project.
Rust is an OSS programming language governed by multiple teams \cite{rust2022}, and has been the `most loved' language for seven years in a row \cite{SO2022survey}.    
Hundreds of companies around the world use Rust in production, replacing critical systems previously written in C/C++ \cite{rust2022}. Since late 2022, Rust can also be used to write Linux kernel components.
One report estimated that 28\% of commits to Rust were by paid contributors \cite{oneil2021coproduction}.
There are considerable concerns about corporate involvement, most notably from Amazon.com, Inc.
In September 2021, Steve Klabnik, a prominent member of Rust's Core Team at the time, tweeted \cite{steve2021rust2}: \textit{``At some point, we all have to have a serious conversation about Amazon's involvement in Rust.''} 
Another member also highlighted a lack of responsiveness to community concerns towards corporate involvement \cite{coreteamtoxic2021}:
\textit{``The core team repeatedly dismissed and maligned several members concerns about the involvement of a company so heavily involved with producing spyware.''}
Shortly after, the full Moderation Team, which is responsible for upholding the code of conduct, resigned on November 22, 2021 \cite{resigned2021,anderson2021}.
The moderation team posted that their resignation is \textit{``in protest of the Core Team placing themselves unaccountable to anyone but themselves''} \cite{resigned2021}.
Around the same time, Klabnik's departure from Rust's Core Team was announced.
The resignation of the moderation team and core developers has gained wide attention in the Rust community \cite{rust_reddit}, which would have led developers in Rust to think more critically about the participation of companies in general, as well as governance issues.

To better understand the differences between paid and volunteer open source contributors, it is important to study the characteristics of these two distinct groups of contributors, which may ultimately lead to better insights to build sustainable open source ecosystems. 
As such, Rust can be considered an intrinsic case study \cite{crowe2011case}, that is, a case study selected on its own merits for its intrinsic features. Thus, the Rust project is a potentially fruitful source to develop a better understanding of the differences and similarities between paid developers and volunteers in the context of a single project.

This study is guided by three main research questions. 
First, we seek to compare paid developers and volunteers in terms of contribution frequency, scale, and tasks. 
Paid developers are more likely used to working on specific tasks that they were assigned than volunteers, who would typically self-select tasks to work on. Further, employed developers typically would work full-time on certain tasks, whereas many volunteers would have normal daytime jobs, and contribute in their spare hours. Thus, we ask: 

\vspace{2pt}
\noindent \textit{\textbf{RQ1:} Do paid developers and volunteers differ in their contribution behavior to the Rust project?}

\vspace{2pt}
Of key importance is OSS projects' continuity and sustainability; long-term contributors play a key role in the long-term health of the project. Thus, our second research question is: 

\vspace{2pt}
\noindent \textit{\textbf{RQ2:} Does being paid or not affect the possibility of a Rust developer becoming a long-term contributor to the Rust project?} 

\vspace{2pt}
Finally, aside from actual contribution, how volunteers perceive paid developers is also important, because volunteers' perceptions of their paid `colleagues' may shape their collaboration with paid developers. Thus, our third research question is: 

\vspace{2pt}
\noindent \textit{\textbf{RQ3:} How do volunteers perceive the participation of paid developers in the Rust project?}

\vspace{2pt} 
We conducted a mixed-methods study to address these questions. 
To answer RQ1 and RQ2, we developed 4 hypotheses, and test these using commit data from Rust's code repository. 
To answer RQ3, we conducted a developer survey to gather volunteer developers' perceptions of paid developers; we focused specifically on the four hypotheses introduced for RQ1 and RQ2.

The results indicate that paid developers and volunteers differ in several ways in how they contribute. For example, core developers who are paid tend to contribute more frequently than core volunteers. 
Further, we found that being paid is positively associated with becoming a long-term contributor in Rust. 
The survey results indicate most volunteers either have prejudices or are unfamiliar with paid developers. 
Understanding the differences and similarities between paid developers and volunteers in OSS projects can aid project leaders in steering their community.

\section{Related Work and Hypotheses}\label{sec: relatedwork}

\subsection{Paid and Volunteer OSS Developers}
A number of studies have previously discussed differences between paid developers and volunteers, with varying ways to distinguish these two groups.
Several early studies established that some OSS developers were paid for their work \cite{Hars02,Hertel2003,Lakhani2005}. These survey studies indicated that considerable numbers of respondents (38\%-55\%) contributed during work time. These studies considered open source development \textit{work} that is effectively paid for by companies who support OSS communities---whether these companies are aware of it or not \cite{Lakhani2005}. Riehle et al.'s 2014 study of project repositories considered contributions made during 9am-5pm during weekdays as `paid work'; based on this heuristic, they estimated that 50\% of OSS work is paid for by companies \cite{Riehle2014Paid}. 
However, a recent study by 
Dias et al. of five company-initiated OSS projects found that most contributions happen between 9am-5pm for \textit{both} paid developers and volunteers \cite{dias2020refactoring}, which casts some doubt on the conclusions by Riehle et al. \cite{Riehle2014Paid}.

The studies cited above do not clearly differentiate between what we label `purposeful' sponsorship through OSS contributions whereby companies purposefully task developers with contributing to OSS projects, in alignment with corporate strategy, and `de facto' sponsorship whereby developers contribute during work time but were not explicitly instructed to do so, or without their managers' approval and awareness.
However, even early studies of OSS highlighted purposeful sponsorship; for example, German reported in 2002 that Red Hat dedicated six full-time developers to the GNOME project
\cite{german2002evolution, german2003gnome}. 
In the 20 years or so that have passed since these early studies, purposeful sponsorship of OSS projects has become much more prevalent \cite[cf.][]{corbet2016linux,linux2017report,oneil2021coproduction}.

Determining whether an OSS developer is paid or a volunteer remains a challenge \cite{claes2018towards,barcomb2018uncovering}. 
Whereas early studies have relied on self-reporting surveys to characterize OSS communities \cite{Hars02,Hertel2003,Lakhani2005},
in more recent years researchers have started to infer this information from software repository data using heuristics, such as the time at which commits are made \cite{Riehle2014Paid}, the email address domain associated with commits \cite{Claes2017abnormal,oneil2021coproduction}, or the 
\texttt{site\_admin} flag that can be set for contributors in GitHub organizations \cite{Dias2018}. Internal developers are those who are paid by the organization who open-sourced the project, whereas external are not employed by that organization. 
In their study on effort estimation in OSS, Robles et al. \cite{robles2022development} distinguished between full-time and non-full-time developers, which implies that those who are full-time are paid by their employer to contribute to OSS projects, whereas those who are non-full-time may be either paid or volunteer.
Barcomb et al. argued that from a community's perspective, volunteers may be indistinguishable from non-volunteers \cite{barcomb2018uncovering}. 

Studying the differences between paid developers and volunteers provides an understanding of these two categories of developers, who may have different reasons to contribute. Dias et al. found that both internal developers and external developers are rather active: internal developers are responsible for ca. 46\% of the pull-requests vs. external developers' ca. 54\% \cite{Dias2018}. Another study 
using the same dataset and heuristic,
investigated differences between employees and volunteers in terms of their contributions and acceptance rates \cite{pinto2018gets}. Volunteers face considerably more rejections (up to 26 times more rejections of contributions) than employees, and have to wait considerably longer than employees (on average 11 vs. 2 days, respectively).
Another study by Dias et al. found that volunteers' contributions focus primarily on refactoring, and that corporate developers (employed by the projects' initiating company) focus more on management (including documentation) \cite{dias2020refactoring}. 

Rather than focusing on \textit{when} contributions are made (work hours vs. spare time), we suggest that a difference between paid developers and volunteers lies in the frequency with which they contribute. Thus, we hypothesize: 

\vspace{2pt}
\noindent \textit{\textbf{H1}: 
Paid developers contribute more frequently than volunteers}. 

\vspace{2pt}
Another way in which paid developers and volunteers might differ is the way they contribute code; Pinto et al. previously observed that contributions by paid developers tend to be larger than those from volunteers \cite{pinto2018gets}. For paid developers, who work on behalf of their companies, achieving their set goals may be more important than concerns such as maintainability of the code \cite{Modularity2014, 4786949}. Merging code changes to an OSS repository is an onerous and time-consuming process for maintainers \cite{10.1145/1368088.1368162}; paid developers may want to reduce the overhead of submitting commits and contribute whole features, or complete tasks with as few commits as possible. Volunteers, on the other hand, may have limited time and do not pursue business-specific requirements to make large code changes. 
For paid developers, these considerations may not exist.  
Thus, we posit:

\vspace{2pt}
\noindent \textit{\textbf{H2}: Paid developers may contribute larger chunks\footnote{Inspired by Kolassa et al. \cite{Commit_Size2013}, we measure chunk as the sum of two variables: lines of code added and lines of code removed in a commit.} of code in commits.} 

\vspace{2pt}
Further, companies have specific business objectives when contributing to OSS, such as integrating their own products \cite{zhang2021companies, zhang2022commercial}, and paid developers are typically assigned particular tasks that are important to their employer's feature roadmap. On the other hand, volunteers tend to contribute to open source projects because of their interest and `passion' \cite{alami2019does,wu07}; while motives will vary, volunteers who contribute to a project will want to see that project succeed and improve over time. This means that rather than delivering business-ready features, we argue they are more likely to make improvements through bug fixes and improving non-functional attributes. 
Thus, we propose:

\vspace{2pt}
\noindent \textit{\textbf{H3}: Contributions by paid developers are more likely features than contributions by volunteers.}

\subsection{Becoming Long-Term Contributors}
A key factor for the sustainability of OSS projects is to attract long-term contributors (LTCs) \cite{zhou2012make, MM15}. 
Open source communities attract a variety of developers with varying motivations to contribute. 
The duration of contributors may vary as well; contributors may contribute only once---so-called one-time contributors \cite{lee2017one,lee2017understanding}. 
Lee et al. \cite{lee2017understanding} found that the main reason for one-time contributors \textit{not} to continue contributing is that they see ``nothing else to contribute.''
Calefato et al. studied open source developers who took a break from contributing \cite{calefato2022will}. Amongst others, they found that all core developers in the 18 projects they studied have taken a break in activity at least once.

Of particular interest to open source projects is, of course, whether developers continue to contribute, i.e., whether they become LTCs.
Several studies have focused on newcomers to open source \cite{casalnuovo2015developer,steinmacher2015social,tan2020first}, developer turnover and retention in open source communities \cite{Zhou2016Inflow,zhang2022turnover,lin2017developer}.
Zhang et al. found a positive association between the diversity of contribution models and the number of volunteers \cite{zhang2021companies} and a negative impact of company domination on the sustainability of OSS projects. 
Valiev et al. \cite{valiev2018ecosystem}, however, found that the involvement of companies has a significant effect on the sustainability of projects in the PyPI ecosystem.

Casalnuovo et al. \cite{casalnuovo2015developer} 
observed that prior experience and a developer's prior social links are important factors in predicting long-term productivity. Prior experience in particular seems important given that without it, social links have a small detrimental effect to cumulative productivity \cite{casalnuovo2015developer}.
Similarly, Zhou and Mockus studied factors that affect the chances that a contributor becomes a LTC \cite{MM15}.
Lin et al. found that developers are more likely to remain active when they start contributing to a project early, when they modify rather than create files, and focus primarily on code rather than documentation \cite{lin2017developer}.

Intrinsic motivations may lead to a long-term bond between volunteers and an OSS project. 
Previous work found that companies may withdraw from OSS projects for various reasons \cite{zhang2022turnover}, for example, when a company changes its business strategy towards an OSS project. In such a case, paid developers may no longer be tasked to work on an OSS project. However, this is not always the case; some paid OSS developers pursue a career as an open source developer and will seek new positions with other companies \cite{schaarschmidt2018company}. 
Thus, we propose: 

\vspace{2pt}
\noindent \textit{\textbf{H4}: Paid developers are less likely to become long-term contributors.}

\subsection{Perceptions on Paid Developers}

Initially, open source software was developed by volunteers; as Raymond characterized it, developers (or `hackers') wrote software to \textit{``scratch an itch''}  \cite{raymond1999cathedral}. As paid developers are now commonly members of open source projects alongside volunteers, it is not surprising that conflicts may arise. While paid developers \textit{may} share a passion and interest for a project, ultimately they have a different ``master to serve.''  An infamous example of this is the case of the Debian project, when a decision to pay two Debian volunteers (the `Dunc-Tank' experiment) led to a considerable uproar in the community \cite{Gerlach2016}, and led some volunteers to reduce their involvement:
 \textit{``Some people who used to do good work reduced their involvement drastically''} \cite{broersma2006}. 
 While there is some work on developers' intentions to accept monetary compensation \cite{krishnamurthy2014acceptance, atiq2016}, the more general question, i.e., how volunteer (unpaid) developers view paid developers, is not well understood.

\section{Study Design}\label{sec: approach}
We conducted a mixed-method study of the Rust project, using both quantitative and qualitative methods \cite{easterbrook2008selecting}. 
Rust was originally created by Mozilla engineer Graydon Hoare in 2006, and is now managed by the Rust Foundation \cite{rust2022}.
We collected and cleaned the commit data of 4,117 Rust contributors (Sec.~\ref{sec: data} and \ref{sec: clean}), conducted comparisons to determine the differences and similarities between paid developers and volunteers (Sec.~\ref{sec: rq1-m}),  
and created a statistical model to determine 
the possibility of becoming long-term contributors (Sec.~\ref{sec: rq2-m}). 
We then conducted a survey among volunteer Rust developers (Sec.~\ref{sec: rq3-m}) to collect their views on paid developers. 
An appendix provides the data, scripts, and other resources  \cite{icse24data}.

\subsection{Data Collection}\label{sec: data}
Rust uses Git as its version control system. We obtained the commit meta-data from GitHub, which hosts the repository of Rust project \cite{Rust}  by querying GitHub's REST API. The time span of the dataset is over 11 years, starting at Rust's creation date (July 7, 2010) until December 16, 2021, containing 114,074 commits. 

\hyphenation{time-stamp}

Each commit includes the name and email of the author, a timestamp, a message description, and the `diffs' (the raw content of changes between different versions of a file).  
Previous research found that some commits are submitted by automated bots rather than human developers \cite{Amreen2020, dey2020exploratory}. We identified four bot accounts, which together submitted 138 commits, based on the patterns identified by previous work \cite{Amreen2020, dey2020exploratory, zhang2021companies, tian2022what}; the list of removed accounts can be found in the online appendix \cite{icse24data}. We also removed rollback/merge commits, leaving a total of 112,969 commits for analysis. Below we describe the procedures for cleaning the data.

\subsection{Data Cleaning} \label{sec: clean}
\subsubsection{Merging Multiple Identities}\label{sec: merge}
Developers may have multiple accounts when contributing to open source projects, each of which may have a different name and email address \cite{bird2006mining, Amreen2020, kouters2012s}. 
To establish an accurate representation of a developer's activity and contributions, it is necessary to merge multiple identities that belong to the same developer.
We addressed this problem by using a rule-based method \cite{zhu2019empirical}, which augments the developer's name and email address, and has been shown to result in a high level of accuracy (with a precision close to 100\%). 
Using this approach on an initial list of 4,673 author identities, 556 identities were merged, resulting in a list of 4,117 distinct authors. To assess the accuracy of this identity merge, we performed a manual verification described in Sec.~\ref{sec: rq3-m}, through which we established the accuracy of developers' identities has a 95\% confidence interval of [0.99,1]. 

\subsubsection{Identifying Paid Developers} \label{sec: affi}

As mentioned earlier, determining whether a developer is paid or a volunteer is not straightforward because developer affiliations are not directly recorded in Git commits \cite{zhang2021companies}, and the Rust community does not have an official record of its contributors' affiliations. We followed approaches used in other studies \cite{Zhou2016Inflow, Claes2017abnormal,oneil2021coproduction}. We first identified a developer's affiliation at the time of each commit they made to Rust by the domain of their email address. More specifically, if a developer uses an email with a free or general provider domain, such as ``gmail.com'', we considered them to be a volunteer. We used a list of free email provider domains maintained in GitHub \cite{brianjones},  which has been verified and used in other studies \cite{valiev2018ecosystem, zhang2020companies}, to identify volunteers in the Rust project. Similarly, we considered that every developer using an email registered at a company or organization domain is a paid developer. For example, if a developer uses an email that ends with ``@mozilla.com'',  they were classified as a Mozilla employee, i.e., a paid developer in Rust. 

While this method works in many cases, this technique is not perfect. Paid developers might use a free email address to submit commits, even when they do so on their employer's behalf---or, indeed, vice versa. To improve the accuracy of developer affiliations, we conducted additional checks by searching the Internet (using ``Rust'' and the developer's name as keywords) and inspected the first 20 results. We analyzed pages from LinkedIn or Rust's official website, or the developer's resum\'{e} if it was accessible, to determine whether a developer is a volunteer or paid to make contributions to Rust. We used this process to manually validate the top 500 developers in our dataset, ordered by the number of commits, who together made 100,136 (87.9\%) commits. 
In only 17 cases (3.4\%), we identified discrepancies, i.e., developers contributing with a corporate email address (e.g. @att.net), while they specified on their CV to be a volunteer contributor to Rust. In those cases, we registered their affiliation as ``volunteer'' and checked all developers with the same domains. Finally, we also performed a manual verification with developers described in Sec.~\ref{sec: rq3-m} and obtained a 94.3\% accuracy. We identified 250 paid developers from 55 companies.

\subsection{Comparing Paid and Volunteer Developers}
\label{sec: rq1-m}

The first analysis sought to determine whether paid and volunteer contributions to Rust differ, considering three metrics: contribution frequency (\textit{H1}), commit size (\textit{H2}), and task categories (\textit{H3}).

\subsubsection{Measuring Contribution Performance} \label{sec: measure}

Contribution frequency was measured by the number of commits submitted by a developer within a fixed time frame. Following previous studies \cite{overney2020not, vasilescu15, valiev2018ecosystem}, we set the time frame to be one month, i.e., if a developer has contributed 100 commits during 3 months, their contribution frequency will be 33.3 (=$\frac{100}{3}$).
The commit size was measured by a widely used metric \cite{vasilescu15, tan2020scaling, zhang2022turnover}: lines of code (LOC) in a commit. For each developer, we took the median of the LOC of all contributed commits to represent their commit size.

To determine the type of work carried out in a commit, we adopted the classification of dos Santos and Figueiredo \cite{Santos2020CommitCU}: `feature,' corresponding to new feature introduction; `corrective,'
related to fault fixing; `perfective,' based on system improvements; and `non-functional,' referring to documentation and non-functional requirements. We used their model, which uses natural language processing to classify the commits in our dataset. While other models exist, we used this model because it has a high F-measure (91\%) and is well documented.  
The output is a list of commit types. The model only assigned the `unknown' label to 553 (0.5\%) commits, most of which have an uninformative message, such as \textit{``Apply suggestions from code review.''} When analyzing developers' task preferences, we did not consider commits that were labeled as `unknown.' We validated the performance of the classification model by randomly selecting 150 commits (error margin: 8\%, confidence level: 95\%) and manually labeling by the first two authors. The results show that 84\% of commits are given the same labels from both the classification model and the manual validation. The high consistency can demonstrate the availability of the classification model we selected.

\subsubsection{Classifying Developer Roles}
The distribution of contributions in  OSS projects frequently follows the Pareto principle \cite{Mockus2002Two, goeminne2011evidence, zhang2021companies}:  
a relatively small group of developers (the core) drives most of the work, while a larger  
group of contributors (the periphery) contributes a considerably smaller amount. While the distribution between these two groups varies, Mockus et al. who first observed this identified a typical 80/20 distribution \cite{Mockus2002Two}. 
Independent of whether developers are paid or volunteers, they may be core, peripheral, or even one-time contributors.  
Instead of comparing paid and volunteer developers at a superficial level, we answer RQ1 at a finer granularity by comparing paid and volunteer developers for each group (core, peripheral, one-time). 

We adopt the widely used core-periphery structure \cite{goeminne2011evidence, valiev2018ecosystem} to classify developers' roles based on the commit-based heuristic following previous work \cite{joblin2017classifying, Coelho2018why}. Core developers are deemed to account for 80\% of commits in Rust. The remaining developers are classified as peripheral developers. We also considered the role of one-time contributors,  
because those one-time contributors may choose to stay on and are essential for the long-term viability of OSS \cite{steinmacher2015social}.  
To avoid duplicate comparisons, we removed one-time contributors from the pool of peripheral developers.

\subsubsection{Comparing Differences} \label{sec: test}
Based on the measures of contribution frequency, commit size, and task categories, we explored the three hypotheses posited above to answer RQ1. 
Given the non-normal distribution of the data, we used the non-parametric Mann–Whitney U test  
\cite{Mann-Whitney} to test for differences between paid developers and volunteers in different role groups (i.e., core, peripheral, and one-time). 
To reduce false discoveries due to multiple hypothesis testing, we adjusted all \textit{p-values} using the Benjamini-Hochberg correction method \cite{benjamini1995controlling}.

We also report the effect size \cite{2012Effect}, which assesses the strength of the relationship between investigated variables. We used the library \texttt{statsmodels} for Python to conduct these statistics. 

\subsection{Modeling Long-Term Contributors}
\label{sec: rq2-m}

Following Zhou et al's \cite{zhou2012make} definition of long-term contributors (LTCs),\footnote{defined as: ``\textit{A participant who stays with the project for at least three years and is productive \cite{zhou2012make}.''}} we characterized a contributor as an LTC if: 
\begin{itemize}
\item they have contributed to Rust for 3 years or more, and
\item they rank in the top 20\% in terms of number of commits in at least three years.\footnote{The Pareto principle (20/80) phenomenon have been frequently encountered in software engineering \cite{Mockus2002Two, zhou2017scalability, zhang2021companies}, and so we deem 20\% a reasonable threshold.}
\end{itemize}
Both requirements should be satisfied. 
We found that 1,508 out of 4,117 developers (36.6\%) joined Rust less than three years at the time of the study and thus we could not assess whether they would become an LTC (based on the definition above); thus, we removed these developers in this analysis.

After determining whether a developer has been an LTC, we fit logistic regression models \cite{2011Logistic} to investigate the association between being paid or not, and the possibility of becoming an LTC, and report the effect size of the independent variable with odds ratio. Previous studies \cite{zhou2012make, MM15} have found that a newcomer's development ability, willingness to participate, and the environment at the time of joining are associated with the likelihood of becoming an LTC. Following previous work \cite{zhou2012make, MM15}, we measured a new contributor’s willingness and ability by the number and types of tasks (e.g., the willingness and ability of fixing bugs are stronger than writing documents \cite{zhou2012make, MM15}). Moreover, we extend the measurement of contributors’ willingness and ability to the commit size (measured by LOC) because submitting large code changes may require huge efforts (what we would call `strong willingness') and can also convey a developer's ability. Since all developers share the same environment (i.e., Rust community), we excluded the environment factor. Besides, we also included gender as one control variable, because existing studies, e.g., \cite{qiu2019going, qiu2023gender}, show that women are at higher risk of leaving an OSS project. We used the NamSor tool \cite{carsenat2019inferring} to automatically infer the genders of contributors based on their names. NamSor is one of the most accurate tools \cite{sebo2021performance}.  
The rest measures are calculated based on the commit data of developers during their first month of participation in Rust, as we sought to investigate whether being paid is linked to the probability for a newly joined contributor to become an LTC.

\subsection{Volunteers' Views on Paid Developers}\label{sec: rq3-m}
The quantitative comparison of paid developers and volunteers presented in previous sections can convey behavioral differences. How OSS volunteers \textit{perceive} paid developers cannot be determined from archival data, but clearly plays a role in whether their collaboration is harmonious or subject to conflict.  
Therefore, we conducted a qualitative study to gain more insight into how volunteers perceive paid developers in Rust, while focusing on the four hypotheses.

In the survey, we asked whether they agreed with the four hypotheses (using a 5-Point Likert Scale, with anchors 1=Strongly disagree and 5=Strongly agree) \cite{joshi2015likert} and their perspectives. Since the contribution performance of paid developers may be strictly determined by their employers or based on their own motivation, we did not fix a corporate option in the survey but included an open-ended question to allow respondents to explain their level of agreement. 

As part of the survey, we also conducted a validation of developer identities. Following previous work \cite{zhang2021companies}, we adopted a less intrusive approach, i.e., for each unique pair of a developer's identity and affiliation, we randomly selected one commit and recorded the affiliation (or labeled as ``volunteer''). We then asked respondents to either confirm or refute that these commits were done by them with the given pair of identity and recorded affiliation.  
While the lead researcher's institution did not require ethics approval for this study, another author is on their institution's ethics review board, and we designed the study following recommended procedures. Cognizant of potential issues, we informed respondents of the aim of this study and confirmed that all responses were treated anonymously. 

We conducted a pilot with five randomly selected developers first. During this process, we communicated with the Executive Director of the Rust Foundation, who offered to help us by calling on developers who received the survey invitation to fill it out. Based on the feedback from the pilot, we added minor changes in some key terms in the hypotheses, such as adding the sentence ``(namely, staying for a relatively long time and making significant contributions)'' to explain what a long-term contributor is. Since the identities of paid developers also need to be checked, we did not exclude them from the survey candidates. We randomly selected 350 objects from developers who were ranked in the top 20\% by their commit count, with an error margin of 5\% and a confidence level of 95\%. We then sent the revised survey to the 350 selected developers; of those, 122 emails were not delivered. 

After two months, we obtained a total of 53 replies, of which five responses came from paid developers (a response rate of 23.2\% ($\frac{53}{350-122}$)). 
Since we sought to understand volunteers' perceptions of paid developers' participation in Rust, only volunteers were asked for their level of agreement on the four hypotheses and further explanations.
The five responses from paid developers were used only to validate their identities and affiliations; one of them indicated a different affiliation that was manually corrected.

We followed the guidelines by Seaman \cite{seaman1999qualitative} to code developers’ explanations of their agreements with the four hypotheses: 1) Two authors first read the original responses in detail. After developing a comprehensive overview of the collected open-ended answers, we examined developers’ responses sentence by sentence and transformed key phrases into short labels as initial codes. 2) Once all response contents were coded, we revisited each code and merged codes where appropriate. For instance, we discovered that ``more working hours'' and ``full-time job'' both suggested that paid developers have more time to make contributions to Rust, then we merged them into one code ``more time''. To minimize the impact of personal bias, a series of face-to-face meetings were conducted (approximately 4 sessions, each lasting 20 minutes or more) to discuss the coding results and resolve any variance in coding. Through this process, we found agreement in the final set of categories of responses listed in Tables 2-5. We provided the coding data and step descriptions in the online appendix \cite{icse24data}.

\section{Results}\label{sec: results}

\subsection{Paid and Volunteer Contributors}
\label{sec: rq1}

\begin{figure}[!h]
\centering
\includegraphics[trim={12mm 20mm 12mm 24mm},clip,width=2.8in]{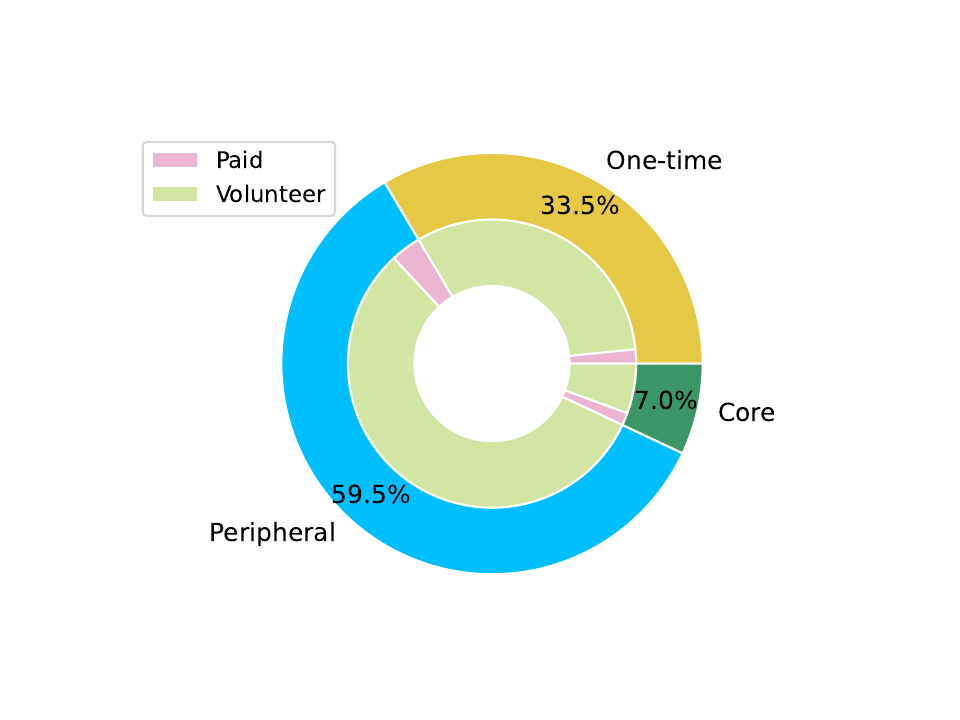}
\caption{Distribution of paid developers and volunteers with one-time, peripheral, and core roles, respectively}
\label{fig: distri}
\end{figure}

To compare paid and volunteer contributors, we first divided developers into three groups (see Sec.~\ref{sec: rq1-m}): one-time, peripheral, and core, and then compared paid developers with volunteers on three metrics in each group: contribution frequency, commit size, and task preference. Figure \ref{fig: distri} shows the distribution of paid and voluntary developers within each of the three groups. Only a small portion of developers that are classified as core, 7.0\% (272), is responsible for 80\% of commits in Rust. The proportion of paid developers in the core group (20.2\%, 55) is higher than in the other two groups (4.8\% in the one-time group and 5.8\% in the peripheral group). 
We analyzed the comparison of contribution between paid and voluntary developers for each group (core, peripheral, one-time) as follows. 

\subsubsection{One-time Developers} 
\begin{figure}[!b]
\centering
\includegraphics[trim={ 0 4mm 0 10mm},clip,width=2.8in]{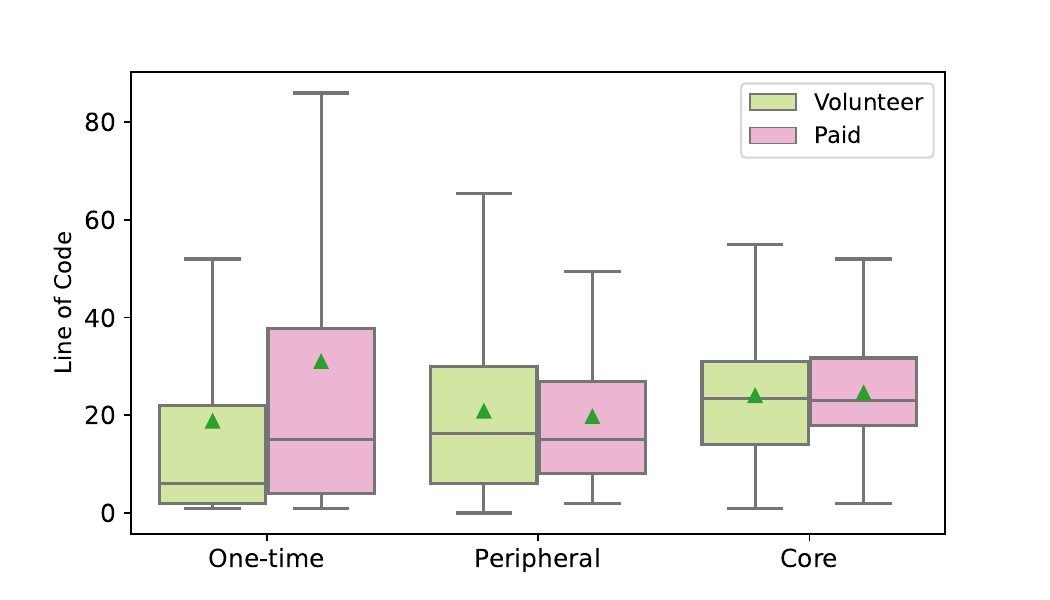}
\caption{LOC distributions of paid developers and volunteers in one-time, peripheral, and core groups}
\label{fig: loc}
\end{figure}

One-time developers have made, by definition, only a single contribution, and thus we cannot compare paid and volunteer developers based on frequency (\textit{H1}), but only on commit size (\textit{H2}) and contribution type (\textit{H3}). We used LOC in each commit as an indicator of the commit size a developer contributes to Rust. 
Figure~\ref{fig: loc} shows that the LOC difference between paid and voluntary developers is greatest for those in the one-time group. Specifically, the median LOC of commits contributed by volunteers is 6, and 15 for paid developers. A Mann–Whitney U test to assess the significance of the difference between one-time paid developers and volunteers in terms of the LOC distributions shows a statistically significant difference (\textit{adjusted p} $=$ .004), although with a small effect size\footnote{effect size $\geq 0.1$ (small) $\geq 0.3$ (medium), and $\geq 0.5$ (large)  
\cite{2012Effect}.} of .25. The results indicate that one-time paid developers tend to contribute larger code changes to OSS projects than one-time volunteers, in support of \textit{\textbf{H2}}.

\begin{figure}[!b]
\centering
\includegraphics[trim={0 0 0 0mm},clip,width=2.8in]{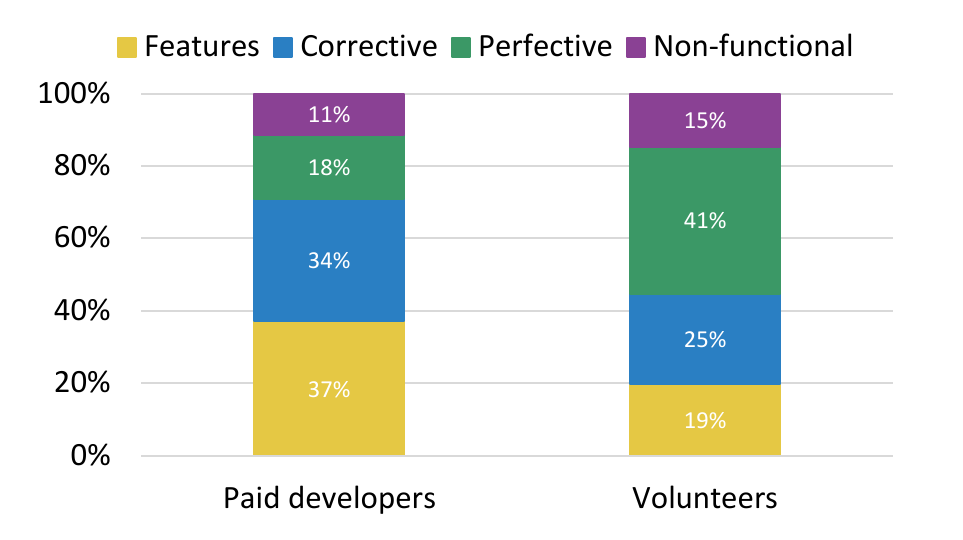}
\caption{Task distributions of one-time  
developers}
\label{fig: ot-task}
\end{figure}

Figure \ref{fig: ot-task} shows the distribution of contribution type (feature, corrective, perfective, non-functional) for one-time paid developers and volunteers. 
Approximately 70\% of one-time commits from paid developers are implementing features and fixing bugs. 
One-time commits from volunteers that add or fix functionality are considerably lower at 44.7\%; further, most commits by volunteers are perfective (40.6\%). More specifically, the feature proportions of the commits contributed by one-time paid developers and volunteers separately are 37.1\% and 19.7\%, respectively. 
These results support \textit{\textbf{H3}}, namely that contributions by paid developers are more likely features than those by volunteers.

\subsubsection{Peripheral Developers}
\begin{figure}[!t]
\centering
\includegraphics[trim={0 4mm 0 10mm},clip,width=2.8in]{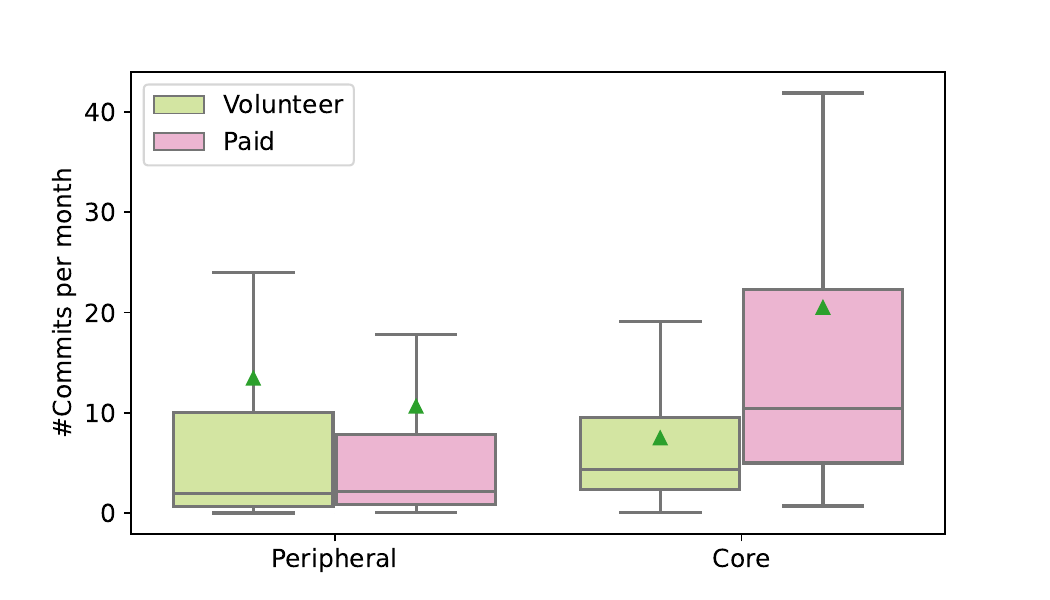}
\caption{Contribution frequency distributions of paid developers and volunteers in peripheral and core groups}
\label{fig: freq}
\end{figure}

As Figure \ref{fig: distri} shows, peripheral developers constitute the largest group in Rust, of which 133 are paid developers, and 2,180 are volunteers. We compare the contribution behavior of paid and voluntary developers in the periphery on contribution frequency, commit size, and task preference. 
The left pair of boxplots in Figure~\ref{fig: freq} shows that the distribution of contribution frequency of volunteers and paid developers is similar.
Specifically, the median frequency for peripheral volunteers is 2 commits per month and 2.2 for peripheral paid developers. 
A Mann–Whitney U test \cite{Mann-Whitney} indicates there is no statistically significant difference (\textit{p} $=$ .96). 
Thus, \textit{\textbf{H1}} is not supported for developers in the periphery.

The second paired box plots in Figure~\ref{fig: loc} show the LOC distribution of peripheral volunteers and paid developers. Specifically, the median LOC of commits contributed is 16 for volunteers, and 15 for paid developers in the peripheral group. Similar to contribution frequency; a Mann–Whitney U test confirmed there is no significant difference (\textit{p} $=$ .85), which means \textit{\textbf{H2}} is not supported for developers in the periphery. 

\begin{figure}[!ht]
\centering
\includegraphics[trim={0 5mm 0 10mm},clip,width=2.8in]{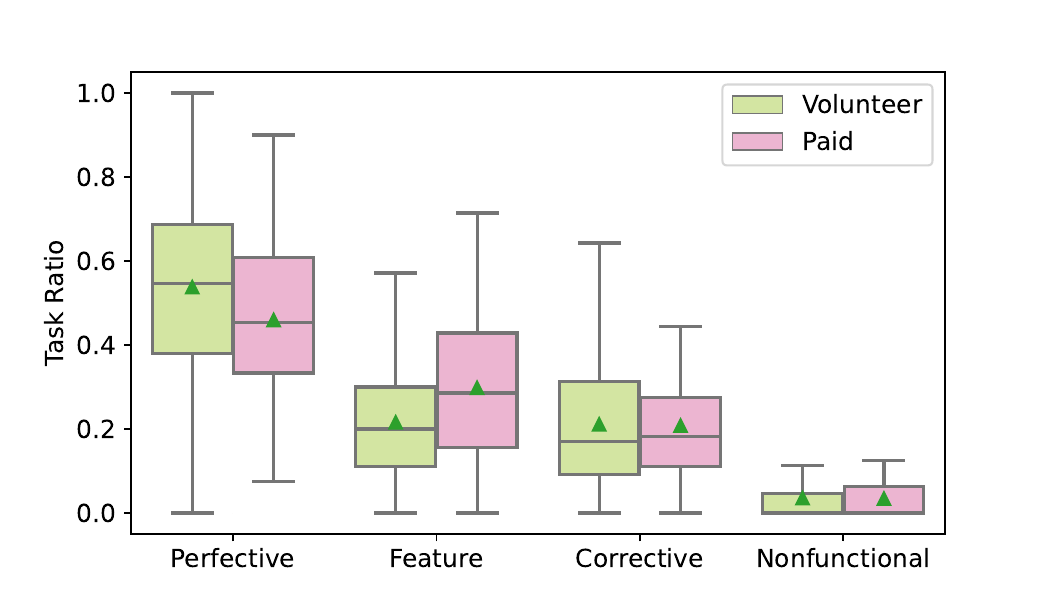}
\caption{Task distributions of peripheral developers} 
\label{fig: task_peri}
\end{figure}

Figure~\ref{fig: task_peri} shows the distribution of four task ratios of voluntary and paid developers in the peripheral group. In the median, 28.6\% of commits contributed by paid developers implement features (20.0\% for volunteers); we observe statistical differences between the percentage distributions of features contributed by volunteers and paid developers (\textit{adjusted p} $=$ .001 and effect size $=$ .26). Further, we can see that perfective commits has the highest ratio in both paid and voluntary developers (18\% vs. 41\%). Specifically, in the median, 45.5\% of the paid developers' commits are perfective (54.5\% for volunteers). A Mann–Whitney U test (\textit{adjusted p} $=$ .014, effect size $=$ .20) indicates that volunteers contribute more commits for perfective tasks than paid developers in the periphery. We found no statistical differences in the corrective and nonfunctional distributions between paid developers and volunteers. These results indicate that, while developers spend most of their time on improving code, paid developers tend to contribute slightly more towards features when compared to volunteers in the peripheral group. These  results support \textit{\textbf{H3}}.

\subsubsection{Core Developers}
Only 7.0\% of developers (i.e., core) account for 80\% of the total commits of Rust. Among the core developers, 55 are paid developers and 217 are volunteers. The right paired box plots in Figure~\ref{fig: freq} present the contribution frequency distribution of core voluntary and paid developers. When compared with peripheral developers, both paid and voluntary developers in the core group have a higher frequency; the median number of contributed commits per month is 4.4 for volunteers and 10.5 for paid developers. 
This difference is significant (\textit{adjusted p} $<$ .001) with a medium effect ($=$ .43), and suggests that paid core developers tend to contribute more frequently than volunteer core developers of Rust. These results lend support to \textit{\textbf{H1}} in the core group. 

The third pair of box plots in Figure~\ref{fig: loc} show the 
LOC distributions of paid and voluntary developers in the core group. Both distributions have similar median and average values. 
For example, core developers, whether they are paid or volunteer, contribute commits with a median of 23 edited lines of code. 
The result (\textit{p}$=$.63) indicates there is no support for \textbf{\textit{H2}} for core developers. 

We also compare the four task type distributions of core contributors. In the median, the most common task is perfective for both paid and volunteers, followed by feature and corrective commits; non-functional task accounts for the smallest proportion of their commits. A Mann–Whitney U test \cite{Mann-Whitney} of feature distributions shows there is no significant difference between paid and voluntary developers. (The appendix presents the statistical results \cite{icse24data}.) Thus, \textit{\textbf{H3}} is not supported for core developers. 

\begin{summarybox}
{
\textbf{Summary for RQ1:} 
Paid developers and volunteers differ in their contribution behavior: paid core developers tend to contribute more frequently to Rust than volunteer core developers;
one-time paid developers tend to contribute bigger commits compared to one-time volunteers; 
and paid peripheral (including one-time) developers tend to focus more on implementing features in comparison to volunteers.
}
\end{summarybox}

\subsection{Becoming Long-Term Contributors}
\label{sec: rq2}

We now address RQ2; in Sec. 2.2 we hypothesized \textit{\textbf{H4}: Paid developers are less likely to become long-term contributors}. To evaluate this, we applied a logistic regression model.

We considered three other factors that could have an impact on the probability of becoming an LTC: new joiners' willingness and ability during the first month of contributing to Rust, and their gender. We used the commit type with the most commits, the number of contributed commits (i.e., the contribution frequency), and the commit size (median LOC in all commits) to measure a developer's willingness and ability. We computed their gender via the NamSor tool.  
We introduced detailed measurements of these factors in Sec.~\ref{sec: rq3-m}. 
Before fitting the attributes in the model, we calculated correlations but found no evidence of collinearity. Developers who joined Rust after 2018-12-16 were excluded because it would be impossible to determine whether or not they are LTCs. The regression is:
\begin{center}
isLTC $\sim$ \emph{Being Paid} + \emph{Contribution Frequency} + \emph{LOC} + \emph{Task Type} + \emph{Gender}
\end{center}

\begin{table}[!b]
\caption{Results of the logistic regression model (\textit{n}=2,609)} 
\centering
\sisetup{
    group-digits=false,
    group-minimum-digits=4,
    table-format=0.2,
    mode=text,
    round-precision=2,
    round-mode=places,
    detect-weight=true, 
    detect-family=true
}
\label{tab: model}
\begin{threeparttable}
\begin{tabular}{p{3.7cm}
                S[table-format=-1.2]
                S[table-format=1.2]
                S[table-format=-1.2]
                S[table-format=1.2]}
\toprule
Variable        & {Coeff.} & {Std.Err} & {z} & {Pr(>|z|} \\ \midrule
(Intercept)            & -4.92 & 0.41 & -11.90 & 0.000 \\
\textbf{Being Paid}    & 1.55  & 0.25 & 6.28  & 0.000 \\
Contribution Frequency & 0.82  & 0.12 & 6.91  & 0.000 \\
LOC                    & 0.17  & 0.11 & 1.57  & 0.116  \\
Corrective             & -0.15 & 0.34 & -0.45 & 0.654 \\
Perfective             & 0.10  & 0.29 & 0.34  & 0.735 \\
Nonfunctional          & 0.01 & 0.58 & 0.01 & 0.99 \\
Female                 & 0.27 & 0.38 & 0.72 & 0.474 \\
\bottomrule
\end{tabular}
\begin{tablenotes}
\footnotesize 
\item[*] LLR p-value = 1.1e-22; Pseudo R-square = 0.13.
\end{tablenotes}
\end{threeparttable}
\end{table}

Table~\ref{tab: model} shows the results of the fitted model. The p-value of this model is 1.1e-22, indicating the regression is statistically significant \cite{LLR}. The positive coefficient and p-value of \emph{Being Paid} indicate that paid developers tend to have a higher probability of becoming long-term contributors when compared with volunteers, contradicting our hypothesis. One reason might be that paid developers have a secure income, a lack of which is a common reason for volunteer turnover in OSS projects \cite{foucault2015impact, 10.1109/SoHeal.2019.00009}. The positive relationship indicates an opportunity to cultivate long-term OSS contributors. Further, future studies aiming to predict LTCs should consider whether contributors are being paid.

As expected, \emph{Contribution Frequency} is statistically significant, suggesting that the intensity of participation in the first month is an indicator of becoming an LTC in Rust. Contributing more commits demonstrates the willingness and ability of newcomers, which are two factors in becoming LTCs, and is consistent with prior work \cite{MM15}. Surprisingly, \emph{Task Type} is not statistically significant, suggesting that in the context of commits, the type of contributions does not influence whether newcomers become LTCs. Similarly, we found that the gender of contributors is not significantly related to whether they will become LTCs. We further checked the distribution of gender in the LTC and non-LTC groups and found that the percentage of LTCs in women and men are comparable (i.e., 4.6\% \textit{vs.} 5.9\%). This indicates that, although women usually have a low percentage in many OSS projects \cite{qiu2023gender}, promoting gender diversity is worthwhile in OSS development (or at least the Rust community).

\begin{summarybox}
{
\textbf{Summary for RQ2:} 
}
Being paid is positively associated with becoming a long-term contributor in the Rust project. 
\end{summarybox}

\subsection{Volunteers' Perceptions of Paid Developers} 
\label{sec: rq3}
To address RQ3, we solicited volunteers' perspectives on paid developers' contributions in terms of the four hypotheses (see Sec.~\ref{sec: relatedwork}). Tables \ref{tab: reason_h1}, \ref{tab: reason_h2}, \ref{tab: reason_h3}, and \ref{tab: reason_h4} present the results. 
We group the reasons for `strongly agree' and `agree' together, and the same for `disagree.' (A single response could include multiple reasons.)

\begin{table}[!b]
\caption{Volunteers' responses to H1}
\vspace{-0.3cm}
\label{tab: reason_h1}
\centering
\begin{tabular}{p{1.3cm}p{5.9cm}r} 
\toprule  
\textmd{Option (n, \%)} & \textmd{Reasons} & {N}  \\ 
\midrule 
\multirow{4}{*}{\shortstack{Agree\\(n=33, 68.8\%)}} & Have more time to contribute & 22 \\
 &  Clear goals reduce time to determine tasks & 8 \\
 &  Obligation forces frequent contributions & 5 \\
 &  No explanation & 7 \\ 
\midrule 
\multirow{3}{*}{\shortstack{Neutral\\(n=10, 20.8\%)}}  & Depend on various factors & 2 \\
 &  Mostly do proprietary projects & 1 \\
 &  No explanation & 7 \\
\midrule 
\multirow{5}{*}{\shortstack{Disagree\\(n=5, 10.4\%)}} & Mostly do proprietary projects & 2 \\
 &  Varies with assignment type & 1 \\
 &  Personal choice & 1 \\
 &  Depend on task difficulty & 1 \\
 &  Paid developers tend to be more senior, do more management & 1 \\
\bottomrule 
\end{tabular}
\end{table}

Table~\ref{tab: reason_h1} shows the synthesized reasons for respondents' level of agreement with \textit{\textbf{H1}}.  
More than half (33, 68.8\%) of respondents (strongly) agreed with \textit{\textbf{H1}}, though reasons varied. Only five respondents (10.4\%) indicated (strong) disagreement towards \textit{\textbf{H1}}, and ten respondents remained neutral. For the responses supporting \textit{\textbf{H1}}, \textit{``having more time''} (including both working hours and spare time) is the most common reason. Further, respondents also pointed out that clear goals and obligations of paid developers force them to contribute more frequently. 

Developers holding opposing or neutral views indicated that task difficulty, personal choice, passions, 
and experiences, also affected developers' contribution frequency. The quantitative analysis showed that being paid only has a significant difference when developers are core members of Rust. This may indicate that only core paid developers can fully work on Rust, while this is not the case for most developers in the periphery. However, 68.8\% of respondents agreed with \textit{\textbf{H1}}, suggesting that most volunteers have a higher expectation of paid developers' contribution frequency.

Table~\ref{tab: reason_h2} shows the synthesized reasons for the respondents' level of agreement with \textit{\textbf{H2}}. Twenty-one (43.8\%) respondents agree (or strongly agree) with \textit{\textbf{H2}}, the same number of respondents hold a neutral stance, and six (12.5\%) respondents disagree with \textit{\textbf{H2}}. The most mentioned supportive reason (n=15) for \textit{\textbf{H2}} is similar as for H1: paid developers have more time to prepare big code changes. On the contrary, one respondent indicated that paid developers tend to make more small changes.  
Six respondents (four `agree' and two `neutral') pointed out that paid developers are assigned with working on specific features, which usually end up with large code dumps. The most common reasons for the `Neutral' and `Disagree' options are the same, i.e., the scale of commits is mainly determined by developers' personal style. The quantitative analysis indicated that only one-time paid developers tend to contribute bigger commits (a median of 15 vs. 6 LOC). This indicates that paid developers who contribute more than once, are suffering from an undeserved stereotype of contributing large chunks of code from over 40\% volunteers. Our results indicate that only one-time paid contributors make considerably larger commits (in the median).

\begin{table}[!t]
\caption{Volunteers' responses to H2}
\vspace{-0.3cm}
\label{tab: reason_h2}
\centering
\begin{tabular}{p{1.3cm}p{5.9cm}r} 
\toprule 
\textmd{Option (n, \%)} & \textmd{Reasons} & N \\ 
\midrule 
\multirow{3}{*}{\shortstack{Agree\\(n=21, 43.8\%)}} & Have more time to prepare big code changes & 15 \\
 & Adding features requires large code changes & 4 \\
 & No explanation & 5\\
\midrule 
\multirow{5}{*}{\shortstack{Neutral\\(n=21, 43.8\%)}}  & Personal choice & 7  \\
 &  Adding features requires large code changes & 2 \\
 &  Have limited time to prepare large changes & 1 \\
 &  Depends on projects & 1  \\  
 &  No explanation & 11 \\
\midrule 
\multirow{4}{*}{\shortstack{Disagree\\(n=6, 12.5\%)}} & Personal choice & 2 \\
 &  Varies with assignment type & 1 \\
 &  Do more small changes & 1 \\
 & No explanation & 2 \\
\bottomrule 
\end{tabular}
\end{table}

\begin{table}[!t]
\caption{Volunteers' responses to H3}
\vspace{-0.3cm}
\label{tab: reason_h3}
\centering
\begin{tabular}{p{1.3cm}p{5.9cm}r} 
\toprule 
\textmd{Option (n, \%)} & \textmd{Reasons} & N \\ 
\midrule 
\multirow{4}{*}{\shortstack{Agree\\(n=14, 29.2\%)}} & Implementing features is more fruitful & 4 \\
 & Company's needs are new features & 4 \\
 & Have more time to implement features & 3 \\
 & No explanation & 4 \\
\midrule 
\multirow{5}{*}{\shortstack{Neutral\\(n=22, 45.8\%)}}  &  Varies with assignment type & 5 \\
 &  Have no preference & 3 \\ 
 &  Personal choice & 3 \\
 &  Have more time to implement features & 1 \\
 & No explanation & 10 \\
\midrule 
\multirow{3}{*}{\shortstack{Disagree\\(n=12, 25.0\%)}} & Do boring work nor interesting features & 4 \\
 &  Varies with assignment type & 4 \\
 &  No explanation & 5 \\ 
\bottomrule 
\end{tabular}
\end{table}

Table~\ref{tab: reason_h3} shows volunteers' perspectives towards \textit{\textbf{H3}}, namely that paid developers focus primarily on adding new features.
Most (22, 45.8\%) respondents remained neutral, 14 (29.2\%) respondents indicated agreement, and 12 (25.0\%)  disagreed with this. For those agreeing, respondents pointed out that implementing features will gain more recognition from an employer than other types of contributions, such as writing documentation, 
and companies' needs from OSS projects are usually adding specific features. 
Features require considerable time to be designed, tested, and implemented. \textit{``Having more time''} is another reason (3 `Agree' and 1 `Neutral' mentioned this). 
These reasons may explain the quantitative results for RQ1, i.e., why one-time and peripheral paid developers are more inclined to contribute features when compared with volunteers. Eight respondents who selected `Neutral' or `Disagree' held the view that, whether or not paid developers prefer implementing features \textit{``varied with assignments''}: some are paid to work on Rust in a way they see fit; others are paid to implement specific features. 
This reason was mentioned in their perspectives of \textit{\textbf{H1}} and \textit{\textbf{H2}}. 
Four volunteers who disagreed with \textit{\textbf{H3}} believe that paid developers tend to \textit{``do boring work nor interesting features,''} which also shows some volunteers' prejudice against paid developers.

\begin{table}[!t]
\caption{Volunteers' responses to H4}
\label{tab: reason_h4}
\centering
\begin{tabular}{p{1.3cm}p{5.9cm}r} 
\toprule 
\textmd{Option (n, \%)} & \textmd{Reasons} & N \\ 
\midrule 
\multirow{4}{*}{\shortstack{Agree\\(n=7, 14.6\%)}} & Lack personal attachment & 3 \\
 & Withdrawal after goal achievement & 2 \\
 & Contribute less if unpaid & 2 \\
 & No explanation & 1 \\
\midrule 
\multirow{5}{*}{\shortstack{Neutral\\(n=26, 54.2\%)}}  & Becoming LTCs is hard for both & 4 \\
 & Vary with assignment type & 3 \\
 & LTCs first then being paid & 2 \\
 & Contribute less if unpaid & 1 \\
  & No explanation & 16 \\
\midrule 
\multirow{3}{*}{\shortstack{Disagree\\(n=15, 31.3\%)}} & Being paid ensures long term & 4 \\
 & LTCs first then being paid & 2 \\
 & No explanation & 9 \\
\bottomrule 
\end{tabular}
\vspace{-5mm}
\end{table}

Table~\ref{tab: reason_h4} shows respondents' categorized perspectives towards their agreements of \textit{\textbf{H4}}, that is,  paid developers are less likely to become long-term contributors. 
Only seven (14.6\%) respondents indicated agreement, and their reasons were in line with our hypothesis: companies may withdraw from OSS projects once their business goal has been achieved, or changes; paid developers may lack personal attachment to the OSS projects and may become less active (or even disappear) when they are no longer paid. More than half of (26, 54.2\%) respondents held a neutral stance. 
Fifteen respondents (31.3\%) disagreed with \textit{\textbf{H4}}: (1) Seven respondents simply indicated this hypothesis contradicted their experience in Rust (or other OSS projects). (2) Four mentioned that being paid ensures long-term contributions because of secure income. (3) Two volunteers explained that developers are usually long-term contributors before being paid by companies. The results of the regression analysis show that being paid is a significantly positive factor in becoming long-term contributors. The reasons for disagreement can be used to explain the modeling results: over half of the neutral perspectives indicate the need to further study the relationship between being paid and developers' long-term participation.

\begin{summarybox}
{
\textbf{Summary for RQ3:} 
Almost 70\% of survey respondents, all of whom are volunteer contributors, agree that paid developers contribute more frequently than volunteers. 
Key reasons are that paid developers have more time, have clear goals, and do so because they are paid. 
Ca. 44\% of respondents believe that paid developers contribute larger commits, while ca. 44\% is unsure, and the remaining 12.5\% disagree. Approx. 70\% volunteers were unsure or disagreed that paid developers focus primarily on adding features. 
Only 14.6\% of volunteers agreed that paid developers are \textit{less} likely to become long-term contributors; ca. 30\% disagreed, and over 54\% were unsure.
}

\end{summarybox}

Table~\ref{tab: summary} presents a summary of the results of the statistical tests and the survey, organized by category of developers. Differences between how paid developers behave confirmed by repository mining and volunteers' perceptions indicate the necessity of informing paid developers' (and companies) contributions in the OSS communities. 

\begin{table}[!t]
\caption{Summary of results}
\vspace{-3mm}
\label{tab: summary}
\centering
\begin{tabular}{p{4mm}p{10mm}p{9mm}p{5mm}p{40mm}} 
\toprule 
Hyp. & One-time & Periphery & Core &Survey findings\\ 
\midrule 
H1 & n/a & \ding{55} & \ding{51} & 33 agreed; 10 were neutral; 5 disagreed. \\ 
H2 & \ding{51} & \ding{55} & \ding{55}  & 21 agreed; 21 were neutral; 6 disagreed. \\
H3 & \ding{51} & \ding{51} & \ding{55} & 14 agreed; 22 were neutral; 12 disagreed.\\
H4 & \multicolumn{3}{c}{Not supported} & 7 agreed; 26 were neutral. 15 disagreed. \\
\bottomrule 
\end{tabular}
\end{table}

\section{Threats to Validity}\label{sec: limitation}

\textit{External validity.} 
This study focused specifically on the Rust project; while the scope of the findings is therefore limited to Rust specifically, future work can test these findings across a large number of projects through a sample study. Besides the three dimensions we have explored, paid developers can also be different from volunteers in other aspects. For instance, are contributions from paid developers more likely to be accepted or rejected in Rust? Future studies could conduct more thorough comparisons between paid developers and volunteers to benefit a well understanding and governance framework of OSS contributors.

A potential threat is the representativeness of our developer sample. Respondents might be unfamiliar with other paid developers. In the survey, we included a neutral option that provides respondents with the opportunity to express a lack of opinion or indifference. This could explain why `Neutral' was the most common answer to H3 and H4. The results from the survey might be biased toward what developers think they know about paid developers. The resignation of the entire moderation team and some core developers leaving Rust attracted considerable attention and some companies' dominating involvement has been one of the most discussed causes \cite{resigned2021, steve2021rust1, steve2021rust2}. Therefore, the issue of paid developers may loom larger in the Rust community than in other OSS projects. The survey received a total of 53 responses (response rate 23.2\%); this number and response rate are similar to other studies of OSS developers \cite{qiu2019going, zhao2017impact, tan2020first}. The survey results complement the quantitative results answering RQ1 and RQ2. However, the survey goal was not to generalize across the Rust project, and so the findings should not be interpreted as such.

\textit{Construct validity}. 
We measured developer contribution through commits, as this represents the key activity in software development.
We acknowledge, as did prior studies \cite{zhang2021companies, zhang2022turnover, Zhou2016Inflow}, that different data sources can be used, including issue reports, code reviews, and online discussions. We decided to use commit data only because the cleaning of data and accurately attributing it to the right group (paid, volunteers) is very time-consuming, as it included a manual check and verification; 
we obtained an accuracy of 94.3\%. Triangulating across other data sources than commit data remains an open challenge for future work. 
The second potential threat lies in distinguishing paid and volunteer developers, which remains an open research challenge. A developer with a public email address could also be paid. On the other hand, using the same heuristic of checking email address domains, a developer who is paid by a company but makes contributions to Rust on their own, may be classified as a paid developer. However, based on the 53 responses, no developers whom we identified as volunteers, indicated they were paid, and only one paid developer indicated their employer is another company. Thus, we deem this threat would have a limited impact, if at all, on the results.

Another decision was our definition of long-term contributors, which required developers to have contributed at least three years to Rust. This definition was based on prior studies \cite{zhou2012make, zhou2015c}. 
While this definition required the exclusion of over 1,500 developers (out of 4,117), we argue that these results are sufficiently representative. Still in answering RQ2, although NamSor leveraged in this study was one the most accurate and convenient methods of gender detection \cite{sebo2021performance}, there are still concerns. For instance, the tool only considers binary genders, and missing information about a country/region can reduce its performance. Future studies can combine the results of multiple gender tools (e.g., Genderize, Gender Predictor, NamSor) to achieve higher reliability. Further, there are other factors, such as social capital \cite{qiu2019going} and the companies of paid developers, which can also affect the possibility of becoming long-term contributors. In this study, we only considered developers' ability, willingness, and gender as control variables, leaving a future research avenue for exploring other factors.

In addition, our survey design did not explicitly ask developers to categorize themselves as a one-time, peripheral, or core contributor. Doing so would constitute a different study, namely how developers perceive their role within a community, which could be linked to important themes such as motivation, sense of belonging, sense of responsibility, and so on. There are potential threats to validity when asking developers directly. First, they may not understand terms and definitions; for example, a one-time contributor may not perceive themselves as a one-time contributor, for example, when they are contributing through other means, or when they have an \textit{intention} to continue to contribute. On one hand, being a one-time, peripheral, or core developer in an OSS project is a significant covariant, i.e., developers in the three groups can exhibit obvious contribution performance. Thus, we deem it necessary to compare paid developers and volunteers in the three groups, which can also construct a better characterization of paid and voluntary contributors. On the other hand, letting developers understand and divide core, peripheral, and one-time contributors is not easy because of different standards. If we mention the three contributor groups in the survey, the questionnaire can become hard to answer. Thus we chose to follow an open question to probe their agreements on the four hypotheses. We argue that if respondents believe being one-time, peripheral, or core developers is an important reason, they will mention them in the following explanation question. 
We chose not to mention the three types of contributors in the questionnaire to minimize the need to explain terms and definitions that could be perceived as overly complex, which in turn could have negative effects on the response rate.

\section{Discussion and Future Work}

This study presents a number of key findings. We find that paid developers' characteristics vary by developer category (one-time, peripheral, core) in Rust.  
Paid developers tend to contribute more frequently than volunteers, but only for core developers. 
The commits contributed by one-time paid developers tend to be larger when compared to one-time volunteers. Except for core developers, paid developers do have a stronger focus on features. Being paid is a positive factor in becoming a long-term contributor. Overall, paid and volunteer developers have their own characteristics, and core paid developers may work beyond their duties.  

\subsection{Governing policies and interface design}

This study presents the first comparative analysis of paid developers and volunteers' contributions within a single project at a fine level of granularity. 
Understanding differences between paid and volunteer contributors can help OSS communities to design better governing policies, and interaction and collaboration interfaces to support the sustainability of OSS projects. For example, OSS communities could conduct real-time measurements of the scale and types of commits contributed by paid developers. When identifying paid developers who are mainly submitting large features, OSS communities could emphasize the need for long-term maintenance and the value of making other types of contributions. 
For volunteers, a lack of sustained time and stable income is perhaps the most pressing barrier that keeps them from becoming long-term contributors. Therefore, OSS communities (and/or hosting platforms) could provide an interface for volunteers who are looking for companies to support them financially (or employ them) to make contributions to projects and promote the interface to companies. Further, differences between paid and volunteer developers observed in this study could be further investigated and included in classifiers for automatically distinguishing paid developers and volunteers with a high degree of accuracy, which remains a non-trivial challenge. 
Future work could further expand the set of dimensions to conduct comparisons between paid and volunteer developers, such as the acceptability of contributions and their preference for different system components or file types. 

\subsection{Volunteers' perspectives on paid developers}

The combined quantitative and survey results suggest that many volunteers might have some `prejudice' against paid developers, such as \textit{``[they] do boring work since [they're] being paid,''} \textit{``[they] rarely care [for] documentation,''} and \textit{``[they] lack personal attachment.''} An unfamiliarity among developers in OSS projects may cause frustration and possibly conflict between paid developers and volunteers, jeopardizing a project's sustainability. To mitigate possible volunteers' biases against paid developers, OSS communities could provide dashboards to transparently visualize companies' contributions in real time. The dimensions studied in this paper (contribution frequency, commit size, task type, likelihood of becoming LTC) provide a set of indicators that could be measured.  

The qualitative results suggest not all paid developers are the same; for example, some are tasked to implement certain required features, and others make contributions as they see fit. 
The latter group may be similar to volunteers but make more frequent contributions, i.e., core paid developers. Future work could verify these hypotheses across a sample of projects and from other perspectives. 
This study suggests that the dichotomous characterization of contributors as either paid or volunteer is too simplistic and does not fully match reality, and that further subcategories could be identified. Future research is necessary to first identify the dimensions that may separate such subgroups of contributors, and second identify, label, and characterize such subgroups. One likely dimension appears to be whether developers have an ambition to pursue a career with the community vs. a career with their company \cite{schaarschmidt2018company}.
Increased awareness of subgroups can support OSS projects in building harmonious relationships between different groups of developers.

\subsection{Improving company participation in OSS}

As suggested by prior work \cite{henkel2006selective, zhang2022commercial} and the survey in this study, there is a perception that companies tend to develop big features in-house that are subsequently contributed to OSS projects.  
Our quantitative analysis confirms this; peripheral paid developers tend to focus more on implementing features than do volunteers. However, such contributions require considerable effort in terms of peer review, adding workload to maintainers of OSS projects who may already face a considerable workload.
Features may directly affect the roadmap of an OSS project.  
Core volunteer developers may perceive these contributions as a sign of detachment and lack of `care' for the project, which could be one reason that corporate participation is disliked by some volunteers \cite{zhang2021companies}. 
Rust, in particular, reportedly had several core developers who left because of the participation of Amazon.com \cite{steve2021rust1, steve2021rust2}. Thus, companies must become more sensitive to the needs and norms of OSS projects, as well as the volunteers in the project, to avoid being perceived as a commercial `parasite.' 
The behavior of even a single paid developer may affect how companies are perceived by volunteers and the wider OSS community.
Companies could use the dashboard mentioned above to inform 
their OSS contribution strategies. As Figure~\ref{fig: distri} shows, the proportion of paid developers in the core group is ca. 20\%, higher than in the other two groups (4.8\% in the one-time group and 5.8\% in the peripheral group), which means companies can be recognized in the Rust project and play a significant role in its development. Assessing whether a company's commercial interests are compatible with an OSS project's roadmap should be the first step before joining. We suggest that hiring volunteers who are already contributing to a project to implement a company's objectives might be a more appropriate and convenient solution because they may be better able to balance the community's long-term concerns and the company's business objectives. 

In sum, it is clear that differences exist across different groups of contributors to an OSS project. This study sought to go beyond previous characterizations of paid vs. volunteer developers, offering a more fine-grained analysis. Given the important role of OSS in today's IT landscape, it is imperative to explore more types of OSS projects, such as single vendor open source projects \cite{riehle2020single}, and understand how evolving OSS communities can remain sustainable; we believe the suggestions for future work outlined above can help with this goal.

\section*{Acknowledgments}
This work is supported by the National Natural Science Foundation of China Grant (62141209, 62202048, 61825201, and 62232003), Science Foundation Ireland (SFI) grant 13/RC/2094-P2 to Lero, the SFI Research Centre for Software and 15/SIRG/3293, and the Open Fund of National Key Laboratory of Parallel and Distributed Computing (PDL).

\balance
\bibliographystyle{ACM-Reference-Format}
\bibliography{ref.bib}
\end{document}